\documentclass[12pt, draftclsnofoot, onecolumn]{IEEEtran}
\usepackage{amsmath,amssymb}
\usepackage{amsfonts}
\usepackage{amsbsy}
\usepackage{graphicx}
\usepackage{caption}

\usepackage{subfigure}
\usepackage{stfloats}
\usepackage{epstopdf}
\usepackage{color}
\usepackage{etoolbox}
\makeatletter
\patchcmd{\@maketitle}
{\addvspace{0.5\baselineskip}\egroup}
{\addvspace{-1\baselineskip}\egroup}
{}
{}
\makeatother
\begin{document}

\title{\huge Outage Performance of $3$D Mobile UAV Caching for Hybrid Satellite-Terrestrial Networks}

\author{\IEEEauthorblockN{Pankaj K. Sharma, Deepika Gupta, and Dong In Kim} 
	\thanks{Pankaj K. Sharma is with the Department of Electronics and Communication Engineering, National Institute of Technology Rourkela, Rourkela 769008, India. Email: sharmap@nitrkl.ac.in.}
	\thanks{Deepika Gupta is with the Department of Electronics and Communication Engineering, Dr S P M International Institute of Information Technology, Naya Raipur, Naya Raipur 493661, India. Email: deepika@iiitnr.edu.in.}
	\thanks{Dong In Kim is with the Department of Electrical and Computer Engineering, Sungkyunkwan University, Suwon, South Korea. Email: dikim@skku.ac.kr.}
}
\maketitle

\begin{abstract}
In this paper, we consider a hybrid satellite-terrestrial network (HSTN) where a multiantenna satellite communicates with a ground user equipment (UE) with the help of multiple cache-enabled amplify-and-forward (AF) three-dimensional ($3$D) mobile unmanned aerial vehicle (UAV) relays. Herein, we employ the two fundamental most popular content (MPC) and uniform content (UC) caching schemes for two types of mobile UAV relays, namely fully $3$D and fixed height. Taking into account the multiantenna satellite links and the random $3$D distances between UAV relays and UE, we analyze the outage probability (OP) of considered system with MPC and UC caching schemes. We further carry out the corresponding asymptotic OP analysis to present the insights on achievable performance gains of two schemes for both types of $3$D mobile UAV relaying. Specifically, we show the following: (a) MPC caching dominates the UC and no caching schemes; (b) fully $3$D mobile UAV relaying outperforms its fixed height counterpart. We finally corroborate the theoretic analysis by simulations.
\end{abstract}

\section{Introduction}
\IEEEPARstart{H}{ybrid} satellite-terrestrial networks (HSTNs) with integrated terrestrial cooperative communications can effectively mitigate the impact of deleterious masking effect in satellite links and thus, are among the promising candidates for next-generation wireless systems \cite{r6}, \cite{r9}. As a result, a handful of works have investigated the performance of general HSTNs by incorporating the fundamental amplify-and-forward (AF)/decode-and-forward (DF) relaying techniques \cite{bhatna}-\cite{sreng}. Also, some works have recently investigated the performance of cognitive HSTNs \cite{pks}, \cite{guo}. In majority of these HSTNs, the satellite and terrestrial link channels are modeled as Shadowed-Rician (SR) and Nakagami-\emph{m} distributed. Despite of substantial efforts made towards improving the performance of HSTNs, these are inevitably affected by the intrinsic issues, namely bandwidth scarcity and latency \cite{bhatna}, \cite{pks}. 

Recently, wireless caching \cite{cch} has been emerged as an effective paradigm where some popular contents are prefetched locally by the network nodes (e.g., relay, etc.) in their installed storage during off-peak hours. Such contents  can be directly fetched at destination through the caching relay node in one hop which alleviates the need of its transmission from source in two hops. It reduces the overall transmission time from source-to-destination by half to tackle the intrinsic spectral efficiency and latency issues. There are two fundamental caching schemes proposed in literature \cite{cch}, namely most popular content (MPC) caching and uniform content (UC) caching. While the MPC caching achieves the largest cooperative diversity gain, the UC caching achieves the largest content diversity gain. Specifically, in the MPC caching scheme, the frequently-demanded (i.e., most popular) content by the user is cached at each network node. Therefore, a cached content can be provided by one of the opportunistically selected network nodes to the user resulting in a cooperative diversity gain without any content diversity. Unlike the MPC caching, in the UC caching, each network node caches different contents uniformly from the available catalogue of data. As a result, the largest content diversity is achievable in the UC caching without cooperative diversity since a unique content can be provided by at most one network node to the user. Although neither of MPC and UC are optimal, these remain fundamental caching schemes for relay networks. A few works \cite{r1}, \cite{r4} have recently applied wireless caching for relay networks. In \cite{r1}, the fundamental MPC and UC caching schemes were studied for AF relay networks. Different from these schemes, a hybrid caching with optimization was studied in \cite{r4}. Counting on numerous advantages, the wireless caching has also been introduced to satellite communications \cite{r7}-\cite{r5}. In \cite{r5}, the MPC and UC caching schemes for HSTNs but for terrestrial Rayleigh fading. 

Furthermore, the unmanned aerial vehicles (UAVs) have recently been considered as promising candidates for future wireless networks. Backed by the low cost, portability, and three-dimensional ($3$D) mobility features, a rotary-wing type UAV is of particular choice for $3$D mobile relaying. A plethora of recent works  \cite{mba}-\cite{pank} have attempted to present stochastic UAV mobility models for the performance analysis of UAV-based wireless networks. For instance, in \cite{mba}, the authors have proposed various UAV mobility models including the random walk (RW) and random waypoint (RWP). In \cite{mna}, a generalized Gauss-Markov process-based UAV mobility model has been proposed for the airborne networks. Further, in \cite{pank1}, \cite{pank}, a mixed-mobility (MM) model was proposed for $3$D UAV movement process. Eventually, based on the MM model, in \cite{r11}, \cite{pank3} and \cite{r10}, the performance of DF and AF $3$D mobile relaying have been investigated for HSTNs, respectively. Note that despite of its importance, the performance of wireless caching for HSTNs with $3$D mobile UAV relaying is entirely unknown even for fundamental MPC and UC schemes. 
          
Motivated by the above, in this paper, we investigate first time the outage probability (OP) of an HSTN that comprises of a multiantenna satellite communicating with a ground user equipment (UE) via multiple cache-enabled AF $3$D mobile UAV relays. We take into account two generic deployment configurations of UAV relays, namely fully $3$D mobile UAV relays and fixed height $3$D mobile UAV relays. Furthermore, as followed in \cite{r1}, \cite{r5}, we consider the popular MPC and UC schemes for the OP as well as corresponding asymptotic OP analyses to assess their cooperative diversity gains.  

\emph{Notations:} $\mathbb{E}[\cdot]$ represents the statistical expectation. $\|\cdot\|$ denotes the Euclidean norm. The acronyms pdf and cdf stand for probability density function $f_X(\cdot)$ and cumulative distribution function $F_X(\cdot)$ of random variable $X$, respectively.   
\section{System Description}\label{sysmod}
\subsection{System Model}
\begin{figure}[!t]
	\centering
	\includegraphics[width=2.8in]{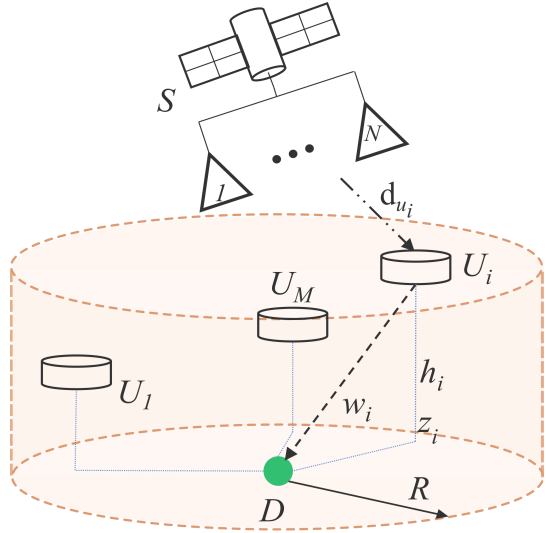}
	\caption{HSTN system model with fully $3$D mobile UAV relays.}
	\label{system}
\end{figure}

As shown in Fig. \ref{system}, we consider an HSTN where a satellite $S$ equipped with $N$ antennas communicates with a single-antenna ground UE $D$ via $M$ cache-enabled single-antenna $3$D UAV relays $U_{i}$, $i\in\{1,...,M\}$. Here, at any time $t$, we assume that the UAVs make $3$D spatial transitions based on MM\footnote{The stochastic MM model may be applied to characterize the random $3$D locations of UAV relays under
some control mechanisms, e.g., altitude control, trajectory control, etc., when the information about their instantaneous locations is unavailable centrally. Note that the stochastic $3$D mobility-based approach is helpful when $3$D spatial point processes lead to a tedious analysis.}model \cite{pank1}, \cite{pank}. The instantaneous altitude of a UAV $U_i$ at time $t$ is denoted by $h_i(t)$ whereas the spatial location is represented as $z_i(t)$. While the UAVs operate in a $3$D cylindrical space of radius $R$ and height $H$ ($H<R$) above the ground plane, the UE $D$ is located at the centre of the base of this cylindrical region. Moreover, we consider that the aforementioned cylindrical region lies beneath the circular spot beam of satellite $S$ centered around the UE $D$. 
All UAVs can update their $3$D locations in discrete time slots. 
The channel vector from $S$ to $U_{i}$ is denoted as $\mathbf{g}_{su_i}\in\mathbb{C}^{1\times N}$ and the channel between $U_i$ and $D$ is denoted as $\textmd{g}_{u_id}$.
All receiving nodes are inflicted by the additive white Gaussian noise (AWGN) with zero mean and variance $\sigma^2$.    

\subsection{Mixed Mobility Model for 3D UAV Movement Process}\label{B}
We consider the multi-parameter MM model \cite{pank1}, \cite{pank} which describes the $3$D mobility of UAVs in a cylinder of height $H$ and radius $R$. The MM model is capable of generating wide range of mobility patterns for UAV relays based on specific setting of underlying parameters. In this model, the UAV makes vertical transitions based on RWP mobility model with random dwell time at each waypoint. at time $t$, the pdf of instantaneous altitude of UAV $h_{i}(t)$ is given by the weighted sum of a static pdf $f^{st}_{h_i}(x|t)$ and a mobility pdf $f^{mo}_{h_i}(x|t)$ as
$f_{h_{i}}(x|t)=p_\mathrm{s}f^{st}_{h_i}(x|t)+(1-p_\mathrm{s})f^{mo}_{h_i}(x|t)$,
where 
$f^{st}_{h_{i}}(x|t)=\frac{1}{H}$
and
$f^{mo}_{h_{i}}(x|t)=-\frac{6x^2}{H^3}+\frac{6x}{H^2}$, for $0 \leq x \leq H,$
with weight $p_\mathrm{s}=\frac{\mathbb{E}[T_\mathrm{s}]}{\mathbb{E}[T_\mathrm{s}]+\mathbb{E}[T_\mathrm{m}]}$ as stay probability at waypoints. 

Meanwhile, in the dwell time, the UAV makes RW in horizontal plane by following $z_i(t+1)=z_{i}(t)+\mathrm{u}_i(t)$ with probability $p_\mathrm{s}$,
where $z_i(t)$ denotes the projection of UAV's location on ground plane and $\mathrm{u}_i(t)$ is the uniform distribution in ball $B(z_i(t),R^{\prime})$ with
$R^\prime$ as the maximum spatial mobility range. Whereas, it follows $z_i(t+1)=z_{i}(t)$ with probability $1-p_\mathrm{s}$. Consequently, the pdf of distance $Z_i(t)=\|z_i(t)\|$ is given by 
$f_{Z_i}(z|t)=\frac{2z}{R^2}, 0\leq z \leq R$.
Various parameters associated with the MM model are described below:
$v_{1,i}(t)\sim[v_{min},v_{max}]$: Uniformly random velocity of vertical transition at waypoints; 	$T_\mathrm{s}\sim[\tau_{min}, \tau_{max}]$ and $\mathbb{E}[T_\mathrm{s}]$: Uniformly random and mean dwell time; $T_\mathrm{m}$ and $\mathbb{E}[T_\mathrm{m}]=\frac{\ln(v_{max}/v_{min})}{v_{max}-v_{min}}\,\frac{H}{3}$: Random and mean vertical movement time; and ${v}_{2,i}(t)=\|z_i(t)-z_{i}(t-1)\|$ and $\mathbb{E}[{v}_{2,i}(t)]=\frac{R^\prime}{1.5}$: Random and mean velocity of horizontal transition at waypoints.  

\subsection{Channel Models}

\subsubsection{Satellite Channel}
For channel vector $\mathbf{g}_{su_{i}}$ whose entries subject to uncorrelated independent and identically distributed (i.i.d.) SR fading, the pdf of $||\mathbf{g}_{su_{i}}||^2$ is given by \cite{pku}
\begin{align}\label{pdfrs}
f_{||\mathbf{g}_{su_{i}}||^2}(x)= \sum_{i_{1}=0}^{m_{su}-1}\cdots \sum_{i_{N}=0}^{m_{su}-1}{\Xi(N)}x^{\gamma-1} \textmd{e}^{-\left({\beta_u-\delta_u}\right)x},
\end{align}
where $\alpha_{u}=(2\flat_{su} m_{su}/(2\flat_{su} m_{su}+\Omega_{su}))^{m_{su}}/2\flat_{su} $, $\beta_{u}=1/2\flat_{su}$, and $\delta_{u}=\Omega_{su}/(2\flat_{su} )(2\flat_{su} m_{su}+\Omega_{su})$,  $\zeta(\kappa)=(-1)^{\kappa}(1-m_{su})_{\kappa}\delta_{u}^{\kappa}/(\kappa!)^{2}$, $(\cdot)_{\kappa}$ is the Pochhammer symbol \cite[p. xliii]{grad}, $\Xi(N)=\alpha^{N}_u\prod_{\kappa=1}^{N}\zeta(i_{\kappa})\prod_{j=1}^{N-1}\mathcal{B}(\sum_{l=1}^{j}i_{l}+j,i_{j+1}+1)$, $\gamma=\sum_{\kappa=1}^{N}i_{\kappa}+N$, and $\mathcal{B}(.,.)$ denotes the Beta function \cite[eq. 8.384.1]{grad}. Further, a free space loss scale factor for satellite links is given as \cite{guo} 
 $\sqrt{\mathcal{L}_{su_i}(t)\vartheta_s\vartheta(\theta_{u_i})}=\sqrt{\frac{\vartheta_s\vartheta(\theta_{u_i})}{\mathcal{K_{B}}\mathcal{TW}}}\left(\frac{c}{4\pi f_\mathrm{c} \mathrm{d}_{u_i}(t)}\right)$,
  where $\mathcal{K_B}=1.38\times10^{-23}$J/K is the Boltzman constant, $\mathcal{T}$ is the receiver noise temperature, $\mathcal{W}$ is the carrier bandwidth, $c$ is the speed of light, $f_\mathrm{c}$ is the carrier frequency, and $\mathrm{d}_{u_i}(t)$ is the distance between $S$ and $U_i$. Here, $\vartheta_s$ denotes the antenna gain at satellite, $\vartheta(\theta_{u_i})$ gives the beam gain of satellite towards $U_i$ which can be expressed as
$\vartheta(\theta_{u_i})=\vartheta_{u_i}\left(\frac{\mathcal{J}_1(\rho_{u_i})}{2\rho_{u_i}}+36\frac{\mathcal{J}_3(\rho_{u_i})}{\rho^3_{u_i}}\right)$,
where $\theta_{u_i}$ is the angular separation of $U_i$ from the satellite beam center, $\vartheta_{u_i}$ is the antenna gain at $U_i$, $\mathcal{J}_\varrho(\cdot)$, $\varrho\in\{1,3\}$ is the Bessel function, and $\rho_{u_i}=2.07123\frac{\sin \theta_{u_i}}{\sin \theta_{{u_i}3\textmd{dB}}}$ with $\theta_{{u_i}3\text{dB}}$ as $3$dB beamwidth.

\subsubsection{UAV Relay-to-Ground Channel}
The terrestrial links between UAV relays $U_i$ and destination $D$ are assumed to follow Nakagami-\emph{m} fading. Thus, the pdf of the channel gains $|g_{u_{i}d}|^2$ belongs to gamma distribution 
\begin{align}\label{gam}
f_{|g_{u_{i}d}|^2}(x)&=\left( \frac{m_{ud}}{\Omega_{ud}}\right)^{m_{ud}} \frac{x^{m_{ud}-1}}{\Gamma(m_{ud})}\,\textmd{e}^{ -\frac{m_{ud}}{\Omega_{ud}} x} ,
\end{align}
where $m_{ud}$ and $\Omega_{ud}$ represent an integer-valued fading severity parameter and average channel power, respectively. 
The instantaneous free-space path loss from UAV $U_i$ to destination $D$ can be given as
$W^{-\alpha}_{id}(t)={\left({h^2_i(t)+Z^2_i(t)}\right)^{-\frac{\alpha}{2}}}$,
where $w_{id}$ is the distance from $U_i$ to $D$ and $\alpha$ is the path loss exponent.

\subsection{Propagation Model}
The satellite $S$ communicates with destination $D$ in two consecutive time phases via variable-gain AF relaying. 

In the first phase, at time $t$, $S$ beamforms its signal $x_{s}(t)$ (satisfying $\mathbb{E}[|x_{s}(t)|^{2}]=1$) to UAV relay $U_i$. Thus, the received signal at $U_i$ can be given by
\begin{align}\label{3}
 y_{u_i}(t)\!&=\!\sqrt{\!P_s\mathcal{L}_{su_i}\!(t)\vartheta_s\vartheta(\theta_{u_i})}\mathbf{g}_{su_i}(t)\mathbf{w}_{su_i}(t)x_{s}(t)\!+\!\nu_{u_i},
	\end{align}
where $\mathbf{w}_{su_i}(t)=\frac{\mathbf{g}_{su_i}(t)}{||\mathbf{g}_{su_i}(t)||}$ is the beamforming weight vector, $\nu_{u_i}$ represents AWGN at $D$ with variance $\sigma^2$. 

In the second phase, at time $t+1$, the UAV relay $U_i$ amplifies and forwards the received signal $y_{u_i}(t)$ to destination $D$ with a gain factor
$G=\sqrt{\frac{1}{P_s\mathcal{L}_{su_i}\!(t)\vartheta_s\vartheta(\theta_{u_i})||\mathbf{g}_{su_i}(t)||^2+\sigma^2}}$.
Thus, the received signal at $D$ can be given as
\begin{align}\label{6}
y_{id}(t+1)&=\sqrt{P_u}W^{-\frac{\alpha}{2}}_{id}(t\!+\!1)Gy_{u_i}(t)\textmd{g}_{u_id}+\nu_{d},
\end{align}
where $\nu_{d}$ is the AWGN at $D$ with variance $\sigma^2$. 
From (\ref{6}), the SNR at $D$ via $U_i$ in the second phase can be calculated as
\begin{align}\label{7}
\Lambda_{id}(t+1)&=\frac{\Lambda_{su_i}(t)\Lambda_{u_id}(t+1)}{\Lambda_{su_i}(t)+\Lambda_{u_id}(t+1)+1},
\end{align}
where 
$\Lambda_{su_i}(t)=\frac{P_s\mathcal{L}_{su_i}(t)\vartheta_s\vartheta(\theta_{u_i})||\mathbf{g}_{su_i}(t)||^2}{\sigma^2}$
and 
$\Lambda_{u_id}(t+1)=\frac{P_u W^{-\alpha}_{id}(t+1)|\textmd{g}_{u_id}(t+1)|^2}{\sigma^2}$. Note that since the distance $\mathrm{d}_{u_i}(t)$ is very large (e.g., $35,786$ Km, for geostationary (GEO) satellite), it is reasonable to assume $\mathrm{d}_{u_i}(t)\approx\mathrm{d}_{u}(t)$, $\mathcal{L}_{su_i}(t)\approx\mathcal{L}_{su}(t)$, $\theta_{u_i}\approx\theta_{u}$, $\rho_{u_i}\approx\rho_{u}$, and $\vartheta_{u_i}\approx\vartheta_{u}$, for all $i$, in this work for subsequent performance analysis.
\subsection{Caching Model and Placement Schemes}
Let us consider a catalogue of $K$ content files of equal size $C$ (in the units of number of files) at UAV relays from which the $k$th file can be requested with popularity defined by the Zipf distribution as    
$f_k=\bigg({k^{\lambda}}{\sum_{k_1=1}^{K}{k_1}^{-\lambda}}\bigg)^{-1}$,
where $\lambda$ denotes the popularity factor. Here, the larger value of $\lambda$ reflects high popularity files. With limited storage capacity at relays $C\ll K$, a fraction of judiciously chosen files can only be stored, i.e., $MC<K$. As described in \cite{r1}, \cite{r5}, we consider the two fundamental yet very popular caching schemes for our HSTN model:
\subsubsection{MPC} If the same content files $C$ are stored at all UAV relays, an arbitrary file with index $k$ can be collaboratively transmitted by a best selected UAV relay according to the criterion ${i}^{\ast}(t)=\displaystyle\arg\max_{i} \Lambda_{u_id}(t), t>1. $
\subsubsection{UC} If different files are stored at each UAV relay, i.e., the file with index $k$ is stored at $U_i$, $(i-1)C+1\leq k\leq iC$. It maximizes the hit probability of finding the required file.

If the requested file is not cached by the UAV relays, then the conventional opportunistic dual-hop relaying is invoked according to ${i}^{\ast}(t)=\displaystyle\arg\max_{i} \Lambda_{id}(t), t>1$. Note that, it may be possible to develop more sophisticated hybrid caching schemes based on a trade-off between MPC and UC schemes which we defer for our future works due to limited space.

%

For the OP analysis in one snapshot under a causal transmission (i.e., $t>1$), the time notation is further relaxed.
\section{Fully $3$D Mobile UAV Relaying}\label{eop}
In this section, we conduct OP analysis for fully $3$D mobile UAV Relaying with no, MPC, and UC caching schemes.
\subsection{OP with No Caching}
With no caching, under i.i.d. SNRs $\Lambda_{id}$, $\forall i$, the OP of HSTN for a rate $\mathcal{R}$ is given by
\begin{align}\label{opt} 
\mathcal{P}^\textmd{NC}_{\textmd{out}}(\mathcal{R})&=[\textmd{Pr}\left[0.5\log_2(1+\Lambda_{id})<\mathcal{R}\right]]^M\\\nonumber
&=[\Psi(\gamma_{\textmd{th1}})]^M,
\end{align}
where the term  $\Psi(\gamma_{\textmd{th1}})$ is evaluated as
\begin{align}\label{8}
\Psi(\gamma_{\textmd{th1}})&=\textmd{Pr}\left[\Lambda_{id}<\gamma_{\textmd{th1}}\right]\\\nonumber
&=1-\int_{0}^{\infty}\left[1-F_{\Lambda_{su_{i}}} \left(\frac{\gamma_{\textmd{th1}}(x+\gamma_{\textmd{th1}}+1)}{x}\right)\right]\\\nonumber
&\times f_{\Lambda_{u_{i}d}}(x+\gamma_{\textmd{th1}})dx,
\end{align}
with $\gamma_{\textmd{th1}}=2^{2\mathcal{R}}-1$. In (\ref{opt}), pre-log factor of $0.5$ reflects two-slot transmissions. In (\ref{8}), the cdf $F_{\Lambda_{su_i}}(x)$  can be evaluated using the pdf in (\ref{pdfrs}) for $\Lambda_{su_i}=\eta_{s}||\mathbf{g}_{su_i}||^{2}$ as
\begin{align}\label{pdflsz}
F_{\Lambda_{su_i}}(x)&=1-\sum_{i_{1}=0}^{m_{su}-1}\cdots \sum_{i_{N}=0}^{m_{su}-1}\frac{\Xi(N)}{(\eta_{s})^{\gamma}}\sum_{p=0}^{\gamma-1} \frac{(\gamma-1)!}{p!}\\\nonumber
&\times \Theta_{u}^{-(\gamma-p)}x^{p}\textmd{e}^{-\Theta_u x}.
\end{align}
where $\Theta_u=\frac{\beta_{u}-\delta_{u}}{\eta_{s}}$ and  $\eta_s=\frac{P_s\mathcal{L}_{su}\vartheta_s\vartheta(\theta_u)}{\sigma^2}$.

Since the mobile UAV relays $U_i$ occupy random locations in $3$D space, the distance $W_{id}$ between $U_i$ and $D$ is random. Its pdf for MM model can be expressed using $p_\mathrm{s}$ as \cite{pank}  
	\begin{align}\label{1u9}
	f_{W_{id}}(w)&=p_\mathrm{s}f^{st}_{W_{id}}(w)+(1-p_\mathrm{s})f^{mo}_{W_{id}}(w),
	\end{align}
	where $f^{st}_{W_{id}}(w)$ is the pdf of $W_{id}$ if $U_i$ makes the horizontal transition and $f^{mo}_{W_{id}}(w)$ is the pdf of $W_{id}$ if $U_i$ makes the vertical transition which are expressed, respectively, as  
	\begin{align}\label{ju11}
	f^{st}_{W_{id}}(w)&=\left\{ \begin{array}{l}
	\frac{2w^{2}}{R^2 H},\textmd{ for }0\leq w<H,\\
	\frac{2w}{R^2},\textmd{ for }H\leq w<R,\\
	\frac{2w}{R^2}-\frac{2w\sqrt{w^2-R^2}}{R^2H},\\
	\qquad\textmd{ for } R\leq w\leq\sqrt{R^2+H^2},
	\end{array}\right.
	\end{align}
	and
	\begin{align}\label{thuo11}
	f^{mo}_{W_{id}}(w)&=\left\{ \begin{array}{l}
	\frac{6w^{3}}{R^2 H^2}-\frac{4w^{4}}{R^2 H^3},\textmd{ for }0\leq w<H,\\
	\frac{2w}{R^2},\textmd{ for }H\leq w<R,\\
	\frac{2w}{R^2}-\frac{6w(w^2-R^2)}{R^2 H^2}+\frac{4w(w^2-R^2)^{\frac{3}{2}}}{R^2 H^3},\\
	\qquad \quad\textmd{ for } R\leq w\leq\sqrt{R^2+H^2}.
	\end{array}\right.
	\end{align}

Further, we derive the pdf $f_{\Lambda_{u_{i}d}}(x)=\frac{d}{dx}F_{\Lambda_{u_{i}d}}(x)$, where 
\begin{align}\label{cdfh}
&F_{\Lambda_{u_{i}d}}(x)	=\textmd{Pr}\left[{\eta_u W^{-\alpha}_{{i}d}|g_{u_{i}d}|^2}<x\right]\\\nonumber	
&{=}\int_{0}^{\sqrt{R^2+H^2}}\frac{\Upsilon\left(m_{ud}, \frac{m_{ud}x}{\Omega_{ud}\eta_u}r^{\alpha}\right)}{\Gamma(m_{ud})} f_{W_{id}}(r)dr,
\end{align}
with $\eta_u=\frac{P_u}{\sigma^2}$. After taking a derivative, we can get the pdf
\begin{align}\label{pdf1}
f_{\Lambda_{u_{i}d}}(x)&=\frac{1}{\Gamma(m_{ud})}\left(\frac{m_{ud}}{\Omega_{ud}\eta_u}\right)^{m_{ud}}x^{m_{ud}-1}\\\nonumber
&\times\int_{0}^{\sqrt{R^2+H^2}}r^{m_{ud}\alpha}\textmd{e}^{-\frac{m_{ud}x}{\Omega_{ud}\eta_u}r^\alpha }\!\!\!f_{W_{id}}(r)dr.
\end{align}
 
Next, we invoke (\ref{pdflsz}) and (\ref{pdf1}) in (\ref{8}) and evaluate the result using \cite[eq. 3.471.9]{grad} to get  
	\begin{align}\label{pouyt1}
	\Psi(\gamma_{\textmd{th1}})&=1-\sum_{i_{1}=0}^{m_{su}-1}\cdots \sum_{i_{N}=0}^{m_{su}-1}\frac{\Xi(N)}{(\eta_{s})^{\gamma}}\sum_{p=0}^{\gamma-1} \frac{(\gamma-1)!}{p!}\\\nonumber
	&\times\sum_{q=0}^{p}\binom{p}{q}\textmd{e}^{-\Theta_u\gamma_{\textmd{th1}}}\sum_{n=0}^{m_{ud}-1}\binom{m_{ud}-1}{n}\\\nonumber
	&\times\frac{2}{\Gamma(m_{ud})}\left(\frac{m_{ud}}{\Omega_{ud}\eta_u}\right)^{m_{ud}-\frac{n-q+1}{2}}\Theta^{\frac{n-q+1}{2}-(\gamma-p)}_u\\\nonumber
	&\times\gamma^{m_{ud}+p-\frac{n+q+1}{2}}_{\textmd{th1}}(\gamma_{\textmd{th1}}+1)^{\frac{n+q+1}{2}}\mathcal{J}_1,
	\end{align}
where the function $\mathcal{J}_1$ is given by
\begin{align}
\mathcal{J}_1&=\int_{0}^{H}J_1(r)J_2(r)dr+\int_{H}^{R}J_1(r)\frac{2r}{R^2}dr\\\nonumber
&+ \int_{R}^{\sqrt{R^2+H^2}}J_1(r)J_3(r)dr,
\end{align}
with $J_1(r)$, $J_2(r)$, and $J_3(r)$ as
\begin{align}
J_1(r)&=r^{\left(m_{ud}-\frac{n-q+1}{2}\right)\alpha}\textmd{e}^{-\frac{m_{ud}\gamma_{\textmd{th1}}r^\alpha}{\Omega_{ud}\eta_u}}\\\nonumber
&\times\mathcal{K}_{n-q+1}\left(2\sqrt{\Theta_u\gamma_{\textmd{th1}}(\gamma_{\textmd{th1}}+1)\frac{m_{ud}\gamma_{\textmd{th1}}r^\alpha}{\Omega_{ud}\eta_u}}\right),
\end{align}
\begin{align}
J_2(r)&=p_\mathrm{s}\frac{2r^{2}}{R^2 H}+(1-p_\mathrm{s})\left(\frac{6r^{3}}{R^2 H^2}-\frac{4r^{4}}{R^2 H^3}\right),
\end{align}
and
\begin{align}
J_3(r)&=p_\mathrm{s}\left(\frac{2r}{R^2}-\frac{2r\sqrt{r^2-R^2}}{R^2H}\right)\\\nonumber&+(1-p_\mathrm{s})\left(\frac{2r}{R^2}-\frac{6r(r^2-R^2)}{R^2 H^2}+\frac{4r(r^2-R^2)^{\frac{3}{2}}}{R^2 H^3}\right),
\end{align}
along with $\mathcal{K}_{v}(\cdot)$ as the $v$th order Bessel function of second kind \cite{grad}. Note that the derived expression can be readily computed numerically by mathematical softwares.

\subsubsection*{Asymptotic OP}
To gain insights on the achievable diversity order of system, we need to simplify the previous OP expression at high SNR ($P_s, P_u\rightarrow \infty$). At high SNR, we invoke $\Lambda_{id}\leq\min(\Lambda_{su_i},\Lambda_{u_id})$ in (\ref{8}) and neglect the resulting higher-order product of cdfs term to achieve 
\begin{align}\label{pouta}
\mathcal{P}^\textmd{NC}_{\textmd{out}}(\mathcal{R})
&\simeq [F_{\Lambda_{su_i}}\left(\gamma_{\textmd{th1}}\right)+F_{\Lambda_{u_id}}\left(\gamma_{\textmd{th1}}\right)]^M.
\end{align}
As followed in \cite{pku}, we can have the simplified cdf $F_{\Lambda_{su_i}}(x)$ under small $x$ as  
\begin{align}\label{dfg}
F_{\Lambda_{su_i}}(x)&\simeq\frac{\alpha^N_u x^N}{N!\eta^N_s}.
\end{align}
 Further, by applying the approximation $\Upsilon(\upsilon,x)\simeq\frac{x^\upsilon}{\upsilon}$, for small $x$, the cdf in (\ref{cdfh}) can be simplified as 
\begin{align}\label{scdu}
F_{\Lambda_{u_id}}(x)&\simeq\frac{1}{m_{ud}!}\left(\frac{m_{ud}x}{\Omega_{ud}\eta_u}\right)^{m_{ud}}\mathcal{J}_2,
\end{align}
where $\mathcal{J}_2$ is 
\begin{align}
\mathcal{J}_2&=\int_{0}^{H}\!r^{m_{ud}\alpha}\!J_2(r)dr+\int_{H}^{R}\frac{2r^{m_{ud}\alpha+1}}{R^2}dr\\\nonumber
&+\int_{R}^{\sqrt{R^2+H^2}}r^{m_{ud}\alpha}J_3(r)dr.
\end{align}

Finally, on substituting the cdf given by (\ref{scdu}) into (\ref{pouta}), the asymptotic OP can be given by 
\begin{align}\label{pouasy}
\mathcal{P}^\textmd{NC}_{\textmd{out}}(\mathcal{R})\!&\simeq\!\left[\frac{\alpha^N_u \gamma^N_{\textmd{th1}}}{N!\eta^N_s}\!+\!\frac{1}{m_{ud}!}\!\left(\frac{m_{ud}\gamma_{\textmd{th1}}}{\Omega_{ud}\eta_u}\right)^{m_{ud}}\mathcal{J}_2\right]^{M}.
\end{align}
\subsection{OP with MPC Caching}
The OP of considered HSTN with MPC can be given as
\begin{align}\label{MPC}
\mathcal{P}^\textmd{MPC}_{\textmd{out}}(\mathcal{R})&=\tilde{\mathcal{P}}^\textmd{MPC}_{\textmd{out}}(\mathcal{R})\sum_{k=1}^{C}f_k+\mathcal{P}^\textmd{NC}_{\textmd{out}}(\mathcal{R})\sum_{k=C+1}^{K}f_k,
\end{align}
where $\tilde{\mathcal{P}}^\textmd{MPC}_{\textmd{out}}(\mathcal{R})$ is defined as (for i.i.d. $\Lambda_{u_{i}d}$)
\begin{align}\label{22}
\tilde{\mathcal{P}}^\textmd{MPC}_{\textmd{out}}(\mathcal{R})&=\textmd{Pr}\left[\log_2(1+\max_i\Lambda_{u_{i}d})<\mathcal{R}\right]\\\nonumber
&=[F_{\Lambda_{u_{i}d}}(\gamma_{\textmd{th2}})]^M
\end{align}
with $\gamma_{\textmd{th2}}=2^{\mathcal{R}}-1$. In (\ref{22}), the pre-log factor of one reflects single-hop transmission. Note that in (\ref{MPC}), the first term corresponds to the case when $D$ successfully fetches the desired locally cached file at UAV relays. This probability of finding the requested file in local cache at a relay is termed as the hit probability \cite{cch}. Whereas, the second term corresponds to the case the desired file is not successfully cached by any relay. This probability of not finding the desired file in local cache is known as the miss probability \cite{cch}. 

\subsubsection*{Asymptotic OP}
Here, at high SNR ($P_s, P_u\rightarrow \infty$), we first simplify the term $\tilde{\mathcal{P}}^\textmd{MPC}_{\textmd{out}}(\mathcal{R})$ using (\ref{scdu}) in (\ref{22}) as 
\begin{align}\label{23}
\tilde{\mathcal{P}}^\textmd{MPC}_{\textmd{out}}(\mathcal{R})&\simeq\left[\frac{1}{m_{ud}!}\left(\frac{m_{ud}\gamma_{\textmd{th2}}}{\Omega_{ud}\eta_u}\right)^{m_{ud}}\mathcal{J}_2\right]^M.
\end{align}
Then, making use of (\ref{pouta}), (\ref{23}) in (\ref{MPC}), the asymptotic OP of MPC can be obtained as
	\begin{align}\label{asympc}
	&\mathcal{P}^\textmd{MPC}_{\textmd{out}}(\mathcal{R})\simeq\left[\frac{1}{m_{ud}!}\left(\frac{m_{ud}\gamma_{\textmd{th2}}}{\Omega_{ud}\eta_u}\right)^{m_{ud}}\mathcal{J}_2\right]^M\sum_{k=1}^{C}f_k\\\nonumber
	&+\left[\frac{\alpha^N_u \gamma^N_{\textmd{th1}}}{N!\eta^N_s}\!+\!\frac{1}{m_{ud}!}\!\left(\frac{m_{ud}\gamma_{\textmd{th1}}}{\Omega_{ud}\eta_u}\right)^{m_{ud}}\mathcal{J}_2\right]^{M}\sum_{k=C+1}^{K}f_k.
	\end{align}
\subsection{OP with UC Caching}
The OP of considered HSTN with UC can be given as
\begin{align}\label{UC}
\mathcal{P}^\textmd{UC}_{\textmd{out}}(\mathcal{R})&=\sum_{i=1}^{M}\tilde{\mathcal{P}}^\textmd{UC}_{\textmd{out},i}(\mathcal{R})\sum_{k=(i-1)C+1}^{iC}f_k\\\nonumber
&+\mathcal{P}^\textmd{NC}_{\textmd{out}}(\mathcal{R})\sum_{k=MC+1}^{K}f_k,
\end{align}
where $\tilde{\mathcal{P}}^\textmd{UC}_{\textmd{out},i}(\mathcal{R})$ is calculated as 
\begin{align}\label{26}
\tilde{\mathcal{P}}^\textmd{UC}_{\textmd{out},i}(\mathcal{R})&=\textmd{Pr}\left[\log_2(1+\Lambda_{u_{i}d})<\mathcal{R}\right]=F_{\Lambda_{u_{i}d}}(\gamma_{\textmd{th2}}),
\end{align}
with $\gamma_{\textmd{th2}}=2^{\mathcal{R}}-1$. Likewise, in (\ref{UC}), the first and second terms represent the hit and miss probabilities, respectively.
\subsubsection*{Asymptotic OP}
The asymptotic OP of UC can be calculated by evaluating (\ref{26}) using (\ref{scdu}) and substituting the result in (\ref{UC}) as 
	\begin{align}\label{asyuc}
&\mathcal{P}^\textmd{UC}_{\textmd{out}}(\mathcal{R})\simeq\sum_{i=1}^{M}\frac{1}{m_{ud}!}\left(\frac{m_{ud}\gamma_{\textmd{th2}}}{\Omega_{ud}\eta_u}\right)^{m_{ud}}\mathcal{J}_2\!\!\sum_{k=(i-1)C+1}^{iC}f_k\\\nonumber
&+\left[\frac{\alpha^N_u \gamma^N_{\textmd{th1}}}{N!\eta^N_s}\!+\!\frac{1}{m_{ud}!}\!\left(\frac{m_{ud}\gamma_{\textmd{th1}}}{\Omega_{ud}\eta_u}\right)^{m_{ud}}\mathcal{J}_2\right]^{M}\sum_{k=MC+1}^{K}f_k.
\end{align}


\section{Fixed Height $3$D Mobile UAV Relaying}
In this section, we perform the OP analysis of MPC and UC schemes for another variant of $3$D mobile UAV relaying where UAV relays are deployed at a fixed altitude $H$ of the cylindrical region of radius $R$ as shown in Fig. \ref{system}. Here, we assume that the UAVs can make transitions in the circular top of the cylindrical region based on the RW mobility model \cite{mba} similar to that described in Section \ref{B}. The resulting steady state distribution of UAV relays remains uniform in the circular disk. For this case, the instantaneous path loss between $U_i$ and $D$ becomes
$W^{-\alpha}_{id}(t)={\left({H^2+Z^2_i(t)}\right)^{-\frac{\alpha}{2}}}$,
where $H$ is the same for all $U_i$. Consequent upon the distribution of $Z_i(t)$ as defined in Section \ref{B}, the pdf of $W_{id}$ in one snapshot can be calculated as  
\begin{align}\label{ju1u1}
f_{W_{id}}(w)&=\left\{ \begin{array}{l}
\frac{2w}{R^2},\textmd{ for }H\leq w\leq\sqrt{R^2+H^2},\\
0,\textmd{ else}.
\end{array}\right.
\end{align}

\subsection{OP with No Caching}\label{4a} 
The OP for this case can be computed similar to (\ref{opt}), i.e., $\mathcal{P}^\textmd{NC}_{\textmd{out}}(\mathcal{R})=[\Psi(\gamma_{\textmd{th1}})]^M$. Further, to compute $\Psi(\gamma_{\textmd{th1}})$ in (\ref{8}), we first need to find the pdf $f_{\Lambda_{u_{i}d}}(x)$ based on the pdf $f_{W_{id}}(w)$ as given by (\ref{ju1u1}) in place of (\ref{1u9}). Then, utilizing the result in (\ref{8}), one can obtain
\begin{align}\label{pouyt11}
	&\Psi(\gamma_{\textmd{th1}})=1-\sum_{i_{1}=0}^{m_{su}-1}\cdots \sum_{i_{N}=0}^{m_{su}-1}\frac{\Xi(N)}{(\eta_{s})^{\gamma}}\\\nonumber
	&\times\sum_{p=0}^{\gamma-1} \frac{(\gamma-1)!}{p!}\sum_{q=0}^{p}\binom{p}{q}\textmd{e}^{-\Theta_u\gamma_{\textmd{th1}}}\sum_{n=0}^{m_{ud}-1}\binom{m_{ud}-1}{n}\\\nonumber
	&\times\frac{2}{\Gamma(m_{ud})}\left(\frac{m_{ud}}{\Omega_{ud}\eta_u}\right)^{m_{ud}-\frac{n-q+1}{2}}\Theta^{\frac{n-q+1}{2}-(\gamma-p)}_u\\\nonumber
	&\times\gamma^{m_{ud}+p-\frac{n+q+1}{2}}_{\textmd{th1}}(\gamma_{\textmd{th1}}+1)^{\frac{n+q+1}{2}}\int_{H}^{\sqrt{R^2+H^2}}J_1(r)\frac{2r}{R^2}dr.
\end{align}

\subsubsection*{Asymptotic OP}
The asymptotic OP can be calculated based on (\ref{pouta}) using the pdf in (\ref{ju1u1}). First, the required asymptotic cdf $F_{\Lambda_{u_id}}(x)$ can be derived based on (\ref{ju1u1}) as  
\begin{align}\label{scdu1}
	F_{\Lambda_{u_id}}(x)&\simeq\frac{1}{m_{ud}!}\left(\frac{m_{ud}x}{\Omega_{ud}\eta_u}\right)^{m_{ud}}\int_{H}^{\sqrt{R^2+H^2}}\frac{2r^{m_{ud}\alpha+1}}{R^2}dr.
\end{align}

Then, substituting the aforementioned cdf along with the cdf $F_{\Lambda_{su_i}}(x)$ given by (\ref{dfg}) into (\ref{pouta}), the asymptotic OP for this case can be determined as
\begin{align}\label{pouasy1}
	&\mathcal{P}^\textmd{NC}_{\textmd{out}}(\mathcal{R})\simeq\left[\frac{\alpha^N_u \gamma^N_{\textmd{th1}}}{N!\eta^N_s}\right.\left.+\!\frac{1}{m_{ud}!}\!\left(\frac{m_{ud}\gamma_{\textmd{th1}}}{\Omega_{ud}\eta_u}\right)^{m_{ud}}\int_{H}^{\sqrt{R^2+H^2}}\frac{2r^{m_{ud}\alpha+1}}{R^2}dr\right]^{M}.
\end{align}
\subsection{OP with MPC Caching}
The OP of MPC can be similarly evaluated as (\ref{MPC}). Here, the term $\mathcal{P}^\textmd{NC}_{\textmd{out}}(\mathcal{R})$ is the same as evaluated previously in Section \ref{4a}. Further, we calculate $\tilde{\mathcal{P}}^\textmd{MPC}_{\textmd{out}}(\mathcal{R})=[F_{\Lambda_{u_{i}d}}(\gamma_{\textmd{th2}})]^M$, where $F_{\Lambda_{u_{i}d}}(\gamma_{\textmd{th2}})$ is obtained by invoking the pdf $f_{W_{id}}(w)$ given by (\ref{ju1u1}) into (\ref{cdfh}) as 
\begin{align}\label{cdfh1}
	&F_{\Lambda_{u_{i}d}}(x)=\int_{H}^{\sqrt{R^2+H^2}}\frac{\Upsilon\left(m_{ud}, \frac{m_{ud}x}{\Omega_{ud}\eta_u}r^{\alpha}\right)}{\Gamma(m_{ud})} \frac{2w}{R^2}dr.
\end{align}
  
\subsubsection*{Asymptotic OP} Also, the asymptotic OP for this case can be determined by substituting the asymptotic expressions of $\mathcal{P}^\textmd{NC}_{\textmd{out}}(\mathcal{R})$ and $\tilde{\mathcal{P}}^\textmd{MPC}_{\textmd{out}}(\mathcal{R})$ in (\ref{MPC}). Note that the asymptotic expression of $\mathcal{P}^\textmd{NC}_{\textmd{out}}(\mathcal{R})$ is already derived as (\ref{pouasy1}). Further, similar to (\ref{23}), we can derive the asymptotic expression of $\tilde{\mathcal{P}}^\textmd{MPC}_{\textmd{out}}(\mathcal{R})$ based on the pdf $f_{W_{id}}(w)$ given by (\ref{ju1u1}) as
\begin{align}\label{2q3}
	&\tilde{\mathcal{P}}^\textmd{MPC}_{\textmd{out}}(\mathcal{R})\simeq\left[\frac{1}{m_{ud}!}\left(\frac{m_{ud}\gamma_{\textmd{th2}}}{\Omega_{ud}\eta_u}\right)^{m_{ud}}\int_{H}^{\sqrt{R^2+H^2}}\frac{2r^{m_{ud}\alpha+1}}{R^2}dr\right]^M.
\end{align}
\subsection{OP with UC Caching}
The OP of UC can be similarly evaluated as (\ref{UC}) where $\tilde{\mathcal{P}}^\textmd{UC}_{\textmd{out},i}(\mathcal{R})=F_{\Lambda_{u_{i}d}}(\gamma_{\textmd{th2}})$ can be computed using (\ref{cdfh1}). Here, the term $\mathcal{P}^\textmd{NC}_{\textmd{out}}(\mathcal{R})$ is the same as derived in Section \ref{4a}.
\subsubsection*{Asymptotic OP} 
Also, the asymptotic OP for this case can be determined by substituting the asymptotic expressions of $\mathcal{P}^\textmd{NC}_{\textmd{out}}(\mathcal{R})$ and $\tilde{\mathcal{P}}^\textmd{UC}_{\textmd{out},i}(\mathcal{R})$ in (\ref{UC}). Note that asymptotic expression of $\mathcal{P}^\textmd{NC}_{\textmd{out}}(\mathcal{R})$ is already derived as (\ref{pouasy1}). Further, we can derive the asymptotic expression of $\tilde{\mathcal{P}}^\textmd{UC}_{\textmd{out},i}(\mathcal{R})$ as
\begin{align}\label{2qq3}
	&\tilde{\mathcal{P}}^\textmd{UC}_{\textmd{out},i}(\mathcal{R})\simeq\frac{1}{m_{ud}!}\left(\frac{m_{ud}\gamma_{\textmd{th2}}}{\Omega_{ud}\eta_u}\right)^{m_{ud}}\int_{H}^{\sqrt{R^2+H^2}}\frac{2r^{m_{ud}\alpha+1}}{R^2}dr.
\end{align}

\emph{\bf{Remark}}: We first compare the asymptotic OP expressions (\ref{pouasy}) and (\ref{asympc}) corresponding to no caching and MPC caching schemes, respectively, for $\eta_s=\eta_u=\eta$, to reveal their achievable diversity order. The diversity order of the system can be inferred as the minimum exponent of $\eta$ in the denominator of these asymptotic OP expressions. We found that both $\mathcal{P}^\textmd{NC}_{\textmd{out}}(\mathcal{R})$ and $\mathcal{P}^\textmd{MPC}_{\textmd{out}}(\mathcal{R})$ varies proportional to $\frac{1}{\eta^{M\min{(N,m_{ud})}}}$ at high SNR. This shows that MPC caching scheme can achieve the full diversity order of that of no caching scheme, i.e., $M\min{(N,m_{ud})}$. However, the MPC caching scheme is expected to offer a relatively higher coding gain over the no caching scheme due to direct single-hop transmission of cached contents. In contrast, the asymptotic OP of UC caching scheme $\mathcal{P}^\textmd{UC}_{\textmd{out}}(\mathcal{R})$ varies proportional to $\frac{1}{\eta^{\min{(MN,m_{ud})}}}$ at high SNR. Hence its achievable diversity order is only $\min{(MN,m_{ud})}$. Note that if $m_{ud}\geq MN$ (e.g., better quality terrestrial links), both MPC and UC schemes have identical diversity order which depends only on the first-hop satellite links. This reveals that under such conditions UC scheme may achieve both the maximum content diversity and cooperative diversity gains. Intuitively, the content diversity gain of UC scheme is $MC$ which is only $C$ in case of MPC scheme.     
\section{Numerical Results}\label{num}
For numerical results, based on \cite{guo}, we set the parameters for satellite as $\mathcal{T}=300$ K, $\mathcal{W}=15$ MHz, $c=3\times10^{8}$ m/s, $\mathrm{d}_{u}=35,786$ Km, $f_\mathrm{c}=2$ GHz, $\vartheta_{u}=4.8$ dB, $\vartheta_{s}=53.45$ dB, $\theta_{u}=0.8^{\circ}$, $\theta_{u 3\textmd{dB}}=0.3^{\circ}$, and $(m_{su}, \flat_{su}, \Omega_{su})=(2, 0.063, 0.0005)$ for heavy shadowing. Based on \cite{pank1}-\cite{pank3}, we set the parameters for UAVs as $v_{1,i}\sim[0.1,30]$ m/s, $v_{2,i}\sim[0,40]$ m/s, $H=80$ m, $R=100$ m, $\Omega_{ud}=1$, and $\alpha=2$. Here, $p_\mathrm{s}=0.5$ is set by adjusting the distribution of $T_{\mathrm{s}}$. We set $\eta_s=\eta_u=\eta$ as SNR and $\mathcal{R}=1$. Further, we perform the simulations on Matlab platform for $10^5$ independent trials. Before every trial, we initially run all UAV relays independently for $10000$ steps based on MM (or RW) model to randomize their locations for fully (or fixed height) $3$D mobile UAV relaying.

\begin{figure}[t]
	\centering
	\includegraphics[width=3.8in]{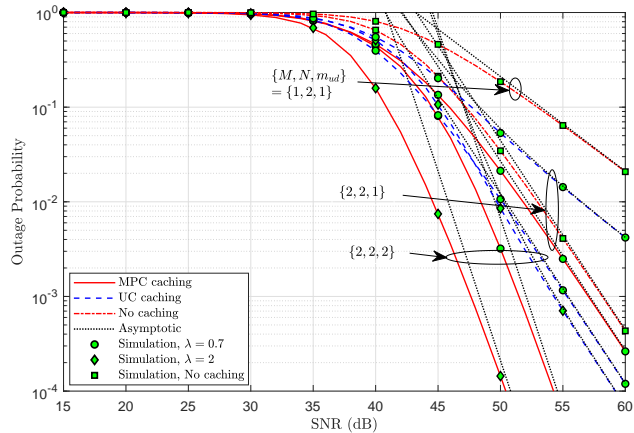}
	\caption{OP of fully $3$D mobile UAV relaying versus SNR.}
	\label{fi11}
\end{figure}

\begin{figure}[t]
	\centering
	\includegraphics[width=3.8in]{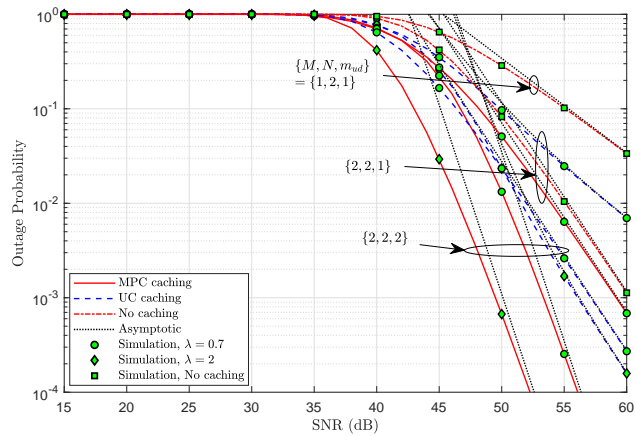}
	\caption{OP of fixed height $3$D mobile UAV relaying versus SNR.}
	\label{fi12}
\end{figure}

Figs. \ref{fi11} and \ref{fi12} plot the OP curves versus SNR for fully and fixed height $3$D mobile UAV relaying, respectively, with no, MPC and UC caching schemes. Apparently, the theoretical, asymptotic, and simulation results are in well agreement with each other. From various curves, we can observe that the MPC caching scheme outperforms significantly the UC and no caching schemes. However, the performance of UC scheme is inferior to the no caching scheme. Firstly, comparing the slopes of curves for $\{M,N,m_{ud}\}=\{2,2,1\}$, we observe that the MPC and no caching schemes have a diversity order of $2$ while UC caching scheme has a diversity order of $1$ (same as no caching with $\{1,2,1\}$). This justifies the fact that the UC caching scheme does not achieve the maximum cooperative diversity gain. Similar observations can be made for $\{M,N,m_{ud}\}=\{2,2,2\}$ where MPC and UC schemes achieve, respectively, the diversity orders of $4$ and $2$. Next, we compare the curves for Zipf parameter $\lambda=0.7$ and $2$ in the set $\{M,N,m_{ud}\}=\{2,2,2\}$. Herein, we find that as $\lambda$ increases from $0.7$ to $2$, the relative performance gain of MPC caching scheme is quite larger as compared to that of the UC caching scheme. However, the UC scheme offers the largest content diversity benefits over its MPC counterpart. Finally, on comparing the set of OP curves of Figs. \ref{fi11} and \ref{fi12}, we conclude that the performance of cache-enabled fully $3$D mobile UAV relaying is remarkably better than that of fixed height one due to lower mean distance between UAV and UE.           
\section{Conclusion}\label{con}
We have evaluated the OP of an HSTN which comprises of a multiantenna satellite, a ground UE and multiple cache-enabled AF $3$D mobile UAV relays. We considered the fundamental MPC and UC caching schemes at UAV relays. In addition, we considered the fully and fixed altitude $3$D mobile UAV relaying for outage performance analysis. We have the following observations: (a) MPC dominates both UC and no caching schemes; (b) fully $3$D mobile UAV relaying achieves better performance gains over the fixed height one.

\end{document}